\documentclass[conference]{IEEEtran}
\IEEEoverridecommandlockouts
\usepackage{cite}
\usepackage{amsmath,amssymb,amsfonts}
\usepackage{graphicx}
\usepackage{textcomp}
\usepackage{xcolor}
\def\BibTeX{{\rm B\kern-.05em{\sc i\kern-.025em b}\kern-.08em
    T\kern-.1667em\lower.7ex\hbox{E}\kern-.125emX}}

\usepackage{glossaries}
\usepackage{bm}
\usepackage{tikz}
\usepackage{dsfont}
\usepackage{threeparttable}
\usepackage{pgfplots}
\pgfplotsset{compat=1.12}
\usepackage{hyperref}

\usepackage{multicol}
\tikzset{near start abs/.style={xshift=1cm}}
\usepackage{amsmath,amsfonts}
\usetikzlibrary{automata}
\usepackage{dirtytalk}
\usetikzlibrary{plotmarks, shapes, positioning, arrows, decorations.markings, fit, calc, patterns, circuits.logic.US}
\usepackage{circuitikz}

\usepackage{algpseudocode}
\usepackage{algorithm}
\usepackage[autostyle]{csquotes}

\begin{document}

\tikzstyle{mux_r}=[trapezium, trapezium stretches = true, rotate=-90, minimum width=1cm, minimum height=0.5cm, draw]
\tikzstyle{branch}=[fill,shape=circle,minimum size=3pt,inner sep=0pt]
\tikzstyle{ckd_node}=[rectangle, minimum width=0.5cm, minimum height=0.5cm, draw]
\tikzstyle{line:thick}=[line width=0.5mm, arrows = {-Latex[width=5pt, length=5pt]}]
\tikzstyle{line:thin}=[line width=0.1mm, arrows = {-Latex[width=5pt, length=5pt]}]
\tikzstyle{line:thick0}=[line width=0.1mm, arrows = {-Latex[width=1pt, length=1pt]}]
\tikzstyle{line:thin0}=[line width=0.1mm, arrows = {-Latex[width=1pt, length=1pt]}]
\tikzstyle{line:thinline}=[line width=0.1mm]
\tikzstyle{node:ALU}=[muxdemux, muxdemux def={NL=2, NR=1, NB=1, w=2, inset w=1, inset Lh=2, inset Rh=0, square pins=1, Lh=6, Rh=3}]
\tikzstyle{line:thick_dot}=[ decorate, decoration={snake, amplitude=.2mm, segment length=3mm}, line width=0.2mm]

\newacronym{alu}{ALU}{arithmetic logic unit}
\newacronym{ascii}{ASCII}{American standard code for information interchange}
\newacronym{asic}{ASIC}{application specific integrated circuit}
\newacronym{at}{A$\times$T}{area $\times$ time}
\newacronym{bia}{BIA}{binary inversion algorithm}
\newacronym{bip}{BIP}{Bitcoin improvement proposal}
\newacronym{ckd}{CKD}{child key derivation}
\newacronym{cntl}{Cntl}{Controller}
\newacronym{cc}{CC}{clock cycle}
\newacronym{ckdf}{CKDF}{child key derivation function}
\newacronym{clb}{CLB}{configurable logic block}
\newacronym{cryp}{crypto}{cryptocurrency}
\newacronym{dpa}{DPA}{differential power analysis}
\newacronym{dsp}{DSP}{digital signal processor}
\newacronym{ec}{EC}{elliptic curve}
\newacronym{ecc}{ECC}{elliptic curve cryptography}
\newacronym{eea}{EEA}{Extended Euclidean algorithm}
\newacronym{fpga}{FPGA}{field programmable gate array}
\newacronym{fsm}{FSM}{finite state machine}
\newacronym{gf}{GF}{Galois field}
\newacronym{hd}{HD}{hierarchically deterministic}
\newacronym{hmac}{HMAC}{hash-based message authentication code}
\newacronym{hsm}{HSM}{hardware security module}
\newacronym{iobs}{IOBs}{input/output buffers}
\newacronym{le}{LE}{logic element}
\newacronym{lsb}{LSB}{least significant bit}
\newacronym{lut}{LUT}{look-up table}
\newacronym{luts}{LUTs}{look up tables}
\newacronym{malu}{MALU}{modular arithmetic logic unit}
\newacronym{mcu}{MCU}{microcontroller unit}
\newacronym{mcus}{MCUs}{microcontroller units}
\newacronym{md}{MD}{Montgomery domain}
\newacronym{msb}{MSB}{most significant bit}
\newacronym{nd}{ND}{non-deterministic}
\newacronym{nist}{NIST}{national institute of standards and technology}
\newacronym{nr}{NR}{not reported}
\newacronym{od}{OD}{original domain}
\newacronym{pa}{ECPA}{elliptic curve point addition}
\newacronym{pbkdf}{PBKDF-2}{password-based key derivation function-2}
\newacronym{pd}{ECPD}{elliptic curve point doubling}
\newacronym{pl}{PL}{programmable logic}
\newacronym{pm}{ECPM}{elliptic curve point multiplication}
\newacronym{prng}{PRNG}{pseudo random number generators}
\newacronym{ps}{PS}{processing system}
\newacronym{qrng}{QRNG}{quantum random number generator}
\newacronym{ram}{RAM}{random access memory}
\newacronym{rng}{RNG}{random number generator}
\newacronym{rtl}{RTL}{register transfer level}
\newacronym{secp}{SECP256K1}{standards for efficient cryptography prime 256 bits Koblitz 1}
\newacronym{sca}{SCA}{side channel analysis}
\newacronym{se}{SE}{secure elements}
\newacronym{sha}{SHA}{secure hash algorithm}
\newacronym{trng}{TRNG}{true random number generator}
\newacronym{tps}{TPS}{transactions per second}
\newacronym{usb}{USB}{universal serial bus}

\title{WiP: Towards a Secure SECP256K1 for Crypto Wallets: Hardware Architecture and Implementation}

\author{Joel Poncha Lemayian, Ghyslain Gagnon, Kaiwen Zhang*, and Pascal Giard \\ 
Department of Electrical Engineering, \'{E}cole de technologie sup\'{e}rieure (\'{E}TS), Montr\'{e}al, Canada\\
*Department of Software Engineering and IT, \'{E}cole de technologie sup\'{e}rieure (\'{E}TS), Montr\'{e}al, Canada\\
Email: joel-poncha.lemayian.1@ens.etsmtl.ca, \{kaiwen.zhang,  pascal.giard\}@etsmtl.ca
}


\maketitle

\begin{abstract}
The SECP256K1 elliptic curve algorithm is fundamental in cryptocurrency wallets for generating secure public keys from private keys, thereby ensuring the protection and ownership of blockchain-based digital assets. However, the literature highlights several successful side-channel attacks on hardware wallets that exploit SECP256K1 to extract private keys. This work proposes a novel hardware architecture for SECP256K1, optimized for side-channel attack resistance and efficient resource utilization. The architecture incorporates complete addition formulas, temporary registers, and parallel processing techniques, making elliptic curve point addition and doubling operations indistinguishable. Implementation results demonstrate an average reduction of 45\% in LUT usage compared to similar works, emphasizing the design’s resource efficiency.
\end{abstract}


\section{Introduction}

\label{sec:intro}

\Gls{cryp} wallets are vital in ensuring the security of digital assets. They generate, store, and manage public and private cryptographic keys. While public keys are openly accessible, the secrecy of private keys is paramount. They serve as proof of ownership, granting full access to associated \gls{cryp} assets \cite{guri2018beatcoin}. Therefore, a compromised private key can result in significant losses. The wallets extensively use cryptographic functions to secure the keys. For example, Ethereum and Bitcoin wallets use the SECP256K1 hash function to generate public keys from private keys. However, studies have demonstrated the extraction of private keys from \gls{cryp} wallets using \glspl{sca} attack, where attackers target the SECKP256K1 \cite{san2019side}. Therefore, this work presents a hardware architecture and implementation for SECP256K1, specifically designed to withstand \gls{sca} attacks.

\subsubsection*{Elliptic Curve Cryptography (ECC):}
SECP256K1 is a specific \gls{ec} in the \gls{ecc} ecosystem. \Gls{ecc} is pivotal in public key cryptography, and it operates on an \gls{ec} defined over a finite \gls{gf} \cite{pirotte2019balancing}. It is used to generate and verify digital signatures and authenticate information. Moreover, \gls{ecc} utilizes shorter cryptographic keys than RSA while providing the same level of security \cite{kabin2020breaking}. Hence, it is attractive for applications with limited resources such as \gls{cryp} hardware wallets.

In \gls{ecc}, two primary finite field types are prevalent: prime fields, denoted as $GF(\mathbb{F}_\texttt{p})$, where $\texttt{p}$ is a large prime number and binary extension fields, denoted as $GF( \mathbb{F}_{2^n})$, where $2^n$ is the field's size, $n$ is a positive integer \cite{pirotte2019balancing}. The \gls{ec} is defined as $y^2 + xy = x^3 + ax + b$, where $x$ and $y$ are coordinates on the \gls{ec}, and $a$ and $b$ are constants that define the curve. After applying a linear transformation to the variables, the \gls{ec} equation is reformulated in the standard short Weierstrass form to $y^2 = x^3 + ax + b$ \cite{kapoor2008elliptic}.

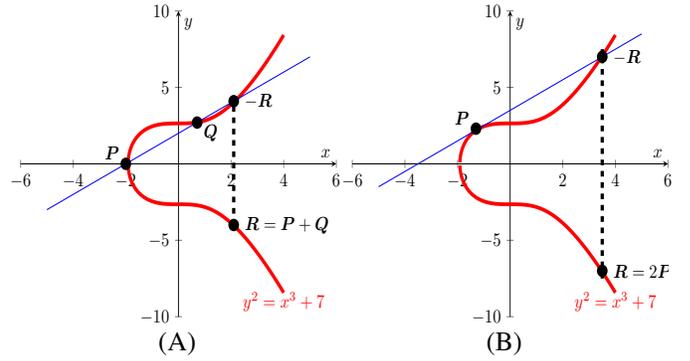
\begin{figure}
    \centering
    \resizebox{0.5\textwidth}{!}{\begin{tikzpicture}[xscale=0.6, yscale=0.7]
    \node[](a) at (3.4,-0.5){(A)};
    \node[](a) at (10.5,-0.5){(B)};
    \begin{axis}[ xmin=-6, xmax=6, ymin=-10, ymax=10,
    axis lines=middle, xlabel=$x$, ylabel=$y$]
        \addplot[line width=2pt, color=red, samples=1001, domain=-4:4]{sqrt(x^3 + 7)};
        \addplot[line width=2pt, color=red, samples=1001, domain=-4:4]{-sqrt(x^3 + 7)};
        \addplot[color=blue, no markers]{x+2};
        \addplot +[color=black, dashed, mark=none, line width=2pt] coordinates {(2.1, -4) (2.1, 4)};
        \node[circle, fill, scale=0.7, color=black,label={[xshift=0.1cm,yshift=0.2cm]left:$\bm{P}$}] at (axis cs:-2,0){};
        \node[circle, fill, scale=0.7, color=black,label={[xshift=-0.1cm,yshift=-0.2cm]right:$\bm{Q}$}] at (axis cs:0.71,2.7){};
        \node[circle, fill, scale=0.7, color=black,label=right:$-\bm{R}$]at (axis cs:2.1,4.1){};
        \node[circle, fill, scale=0.7, color=black,label={right:$\bm{R}=\bm{P}+\bm{Q}$}] at (axis cs:2.1,-4){};
        \node[text=red](eq) at (4,-9){$y^2=x^3+7$};
    \end{axis}
    
 \begin{scope}[xshift=7.2cm]
    \begin{axis}[ xmin=-6, xmax=6, ymin=-10, ymax=10,
    axis lines=middle, xlabel=$x$, ylabel=$y$]
        \addplot[line width=2pt, color=red, samples=1001, domain=-4:4]{sqrt(x^3 + 7)};
        \addplot[line width=2pt, color=red, samples=1001, domain=-4:4]{-sqrt(x^3 + 7)};
        \addplot[color=blue, no markers]{x+3.5};
        \addplot +[color=black, dashed, mark=none, line width=2pt] coordinates {(3.5, -7.5) (3.5, 7.5)};
        \node[circle, fill, scale=0.7, color=black,label={[xshift=0.1cm,yshift=0.2cm]left:$\bm{P}$}] at (axis cs:-1.3,2.3){};
        \node[circle, fill, scale=0.7, color=black,label=right:$-\bm{R}$]at (axis cs:3.5,7){};
        \node[circle, fill, scale=0.7, color=black,label={right:$\bm{R}=2\bm{P}$}] at (axis cs:3.5,-7){};
        \node[text=red](eq) at (4,-9){$y^2=x^3+7$};
    \end{axis} 
 \end{scope}
\end{tikzpicture}}
    \caption{\gls{pa} (A) and \gls{pd} (B) on \gls{ec}. Adapted from \cite{kapoor2008elliptic}.}
    \label{fig:papd}
\end{figure}

Three main processes are carried out on the \gls{ec} to compute public keys. These are \gls{pa}, \gls{pd}, and \gls{pm}\cite{panchbhai2015implementation}. \autoref{fig:papd} (A) shows the \gls{pa} process defined by $\bm{R} = \bm{P} + \bm{Q}$. The figure illustrates the addition of two points, $\bm{P}=(x_0, y_0)$ and $\bm{Q}=(x_1, y_1)$ on the \gls{ec}. As shown, adding $\bm{P}$ and $\bm{Q}$ is equivalent to drawing a straight line through points $\bm{P}$ and $\bm{Q}$, which intersect with the curve at point $\bm{-R}$. Reflecting the point of intersection by the x-axis results in point $\bm{R}$, the \gls{pa} solution.

Conversely, \gls{pd} is the addition of a point on the \gls{ec} by itself. \autoref{fig:papd} (B) depicts the \gls{pd} process defined by $\bm{R} = 2\bm{P}$. The figure illustrates the addition of a point, $\bm{P}=(x_0, y_0)$ with itself on the \gls{ec}. As shown, doubling $\bm{P}$ is equivalent to drawing a line tangent to the \gls{ec} on point $\bm{P}$, which intersects with the curve at point $\bm{-R}$. Reflecting the point of intersection by the x-axis results in point $\bm{R}$, the \gls{pd} solution.

The scalar \gls{pm} process uses the \gls{pa} and the \gls{pd} inside an algorithm, for instance, inside the Montgomery Ladder algorithm \cite{pirotte2019balancing}. \Gls{pm} is the main process used to compute a public key given a private key, and it is defined as $\bm{R} = k\cdot\bm{P}$. It is the sum of $k$ copies of $\bm{P}$, such that:
\begin{equation}
\bm{R} = k\cdot\bm{P} = \sum_{i=1}^{k}\bm{P},
\label{eq_pp}
\end{equation}
where $\bm{R}$ and $\bm{P}$ are points on the curve and $k$ is a long binary number. SECP256K1 is an \gls{ec} whose $a$ and $b$ parameters are 0 and 7, respectively \cite{renes2016complete}. Moreover, \autoref{table:tab_secp} shows other curve parameters. $G$ is the generator point. It is a specific point on the \gls{ec} used as the basis for generating all other points on the curve through \gls{pm}. $G$ is formulated as $G$ = x04 $\parallel$ x $\parallel$ y, where, x04 is a format identifier and x and y are coordinates from a point on the \gls{ec}. 

\begin{table}[h]
\centering
\caption{Parameters for the SECP256K1 \gls{ec}}
\begin{tabular}{p{8.5cm}} 
 \hline
    \texttt{p} = 0x FFFFFFFF FFFFFFFF FFFFFFFF FFFFFFFF FFFFFFFF\\ 
    \hspace{0.85cm}FFFFFFFF FFFFFFFE FFFFFC2F \\
    $G$ = 0x 04 79BE667E F9DCBBAC 55A06295 CE870B07 029BFCDB \\
    \hspace{0.85cm}2DCE28D9 59F2815B 16F81798 483ADA77 26A3C465\\
    \hspace{0.85cm}5DA4FBFC 0E1108A8 FD17B448 A6855419 9C47D08F\\
    \hspace{0.85cm}FB10D4B8\\   
 \hline
\end{tabular}
\label{table:tab_secp}
\end{table}

\subsubsection*{\gls{sca} on SECP256K1:}
In protocols that utilize \gls{ecc}, $k$ is usually considered a private key. Hence, a successful attack correctly derives $k$ via unauthorized means. For example, an \gls{sca} attack can exploit the current drawn or electromagnetic waves emitted by an \gls{ecc} device while processing $k$. The attack relies on the variations in power consumption when bit value 1 or 0 of $k$ is being processed (i.e. $k_i$ where $i$ is the index). 

 The Montgomery Ladder algorithm depicted in \autoref{alg:mont_orig} is popularly used to calculate \gls{pm} \cite{pirotte2019balancing}. The algorithm details the bitwise processing of the secret key $k$ from \gls{msb} to \gls{lsb}. Montgomery Larder is balanced, meaning that the sequence of mathematical operations is independent of the private key. Hence, the literature considers the algorithm safe against simple \gls{sca} attacks \cite{kabin2020breaking}. 

 Nevertheless, the algorithm contains inconsistencies that may make it susceptible to \gls{dpa} and timing attacks. The \gls{pd} in each branch of the \textit{if} statement is performed on different registers. When $m_i$ is 1 \gls{pd} is performed on $\bm{R_1}$. Conversely, when $m_i$ is 0 \gls{pd} is performed on $\bm{R_0}$. These differences create a power consumption pattern or execution time discrepancy, where different registers, memory locations, and data paths are utilized. Such patterns can be exploited by attackers to extract $m$ \cite{kabin2020breaking}. Moreover, the conventional Weierstrass \gls{ec} addition operation contains branching when computing \gls{pa}, \gls{pd}, or a point at infinity. The branching can cause timing variability which can be exploited to reveal the secret key \cite{renes2016complete}.
 
 Various works in literature have proposed ways to protect the Montgomery larder algorithm against \gls{sca} attacks. For example, the work in \cite{pirotte2018design} proposed a method to randomize the sequence of writing $\bm{Q_0}$ and $\bm{Q_1}$ inside the loop. However, branches still access different registers, making the risk of \gls{sca} attacks persistent. 
 
 \begin{algorithm}
\caption{Montgomery Ladder. Adapted from \cite{montgomery1987speeding}}
 \label{alg:mont_orig}
\begin{algorithmic}
    \renewcommand{\algorithmicrequire}{\textbf{Input:}}
    \renewcommand{\algorithmicensure}{\textbf{Output:}}
    \Require $P$, $k=k_l k_{l-1} \dots k_0$, with $k_l = 1$
    \State $Q_0 \gets P$; $Q_1 \gets 2P$
    \For {$i = (l - 2)$ \textbf{downto} $0$}
        \If {$k_i = 1$} \hspace{1em} $Q_0 \gets Q_0 + Q_1$; $Q_1 \gets 2Q_1$
        \Else \hspace{5.4em} $Q_1 \gets Q_0 + Q_1$; $Q_0 \gets 2Q_0$ \EndIf
     \EndFor \label{montg:end-for}
    \Ensure  $Q = kP$
\end{algorithmic}
\end{algorithm}


\begin{algorithm}
 \caption{Point addition equations. Taken from \cite{renes2016complete}}
 \label{alg:eqs}
 \begin{algorithmic}[1]\small
 \renewcommand{\algorithmicrequire}{\textbf{Input:}}
 \renewcommand{\algorithmicensure}{\textbf{Output:}}
 \Require $\bm{P} = (X_1, Y_1, Z_1), \bm{Q} = (X_2, Y_2, Z_2)$ on $E : Y^2Z = X^3 + bZ^3$ and $b_3 = 3 \cdot b.$
 \Ensure  $(X_3, Y_3, Z_3) = \bm{P} + \bm{Q}$;
  \begin{multicols}{3}
    \Statex $t_0 \leftarrow X_1 \cdot X_2$
    \Statex $t_1 \leftarrow Y_1 \cdot Y_2$
    \Statex $t_2 \leftarrow  Z_1 \cdot Z_2$
    \Statex $t_3 \leftarrow  X_1 + Y_1$
    \Statex $t_4 \leftarrow X_2 + Y_2$
    \Statex $t_3 \leftarrow t_3 \cdot t_4$
    \Statex $t_4 \leftarrow t_0 + t_1$
    \Statex $t_3 \leftarrow t_3 - t_4$
    \Statex $t_4 \leftarrow Y_1 + Z_1$
    \Statex $X_3 \leftarrow Y_2 + Z_2$ 
    \Statex $t_4 \leftarrow t_4 \cdot X_3$ 
    \Statex $X_3 \leftarrow t_1 + t_2$
    \Statex $t_4 \leftarrow t_4 - X_3$  
    \Statex $X_3 \leftarrow X_1 + Z_1$ 
    \Statex $Y_3 \leftarrow X_2 + Z_2$
    \Statex $X_3 \leftarrow X_3 \cdot Y_3$  
    \Statex $Y_3 \leftarrow t_0 + t_2$ 
    \Statex $Y_3 \leftarrow X_3 - Y_3$
    \Statex $X_3 \leftarrow t_0 + t_0$  
    \Statex $t_0 \leftarrow X_3 + t_0$ 
    \Statex $t_2 \leftarrow b_3 \cdot t_2$
    \Statex $Z_3 \leftarrow t_1 + t_2$  
    \Statex $t_1 \leftarrow t_1 - t_2$ 
    \Statex $Y_3 \leftarrow b_3 \cdot Y_3$
    \Statex $X_3 \leftarrow t_4 \cdot Y_3$  
    \Statex $t_2 \leftarrow t_3 \cdot t_1$ 
    \Statex $X_3 \leftarrow t_2 - X_3$
    \Statex $Y_3 \leftarrow Y_3 \cdot t_0$  
    \Statex $t_1 \leftarrow t_1 \cdot Z_3$ 
    \Statex $Y_3 \leftarrow t_1 + Y_3$
    \Statex $t_0 \leftarrow t_0 \cdot t_3$  
    \Statex $Z_3 \leftarrow Z_3 \cdot t_4$ 
    \Statex $Z_3 \leftarrow Z_3 + t_0$
\end{multicols}
 \end{algorithmic}
 \end{algorithm}

\subsubsection*{Contribution and Organization:}
This work proposes a hardware architecture and implementation of a SECP256K1 module for \gls{cryp} wallets that are secure against \gls{sca} attacks while utilizing minimum resources hence adhering to the industry standard for small, compact, and portable wallets. The module employs temporary registers and parallel processing to prevent variations during \gls{pm}. Moreover, the complete addition formulas are used to prevent timing variability \cite{renes2016complete}. The remainder of this work is structured as follows. \autoref{sec:hw_arch} describes the hardware architecture of the proposed SECP256K1 function. \autoref{sec:fpga_imp} discusses the implementation results and \autoref{sec:concl} provides the conclusion.

\section{Hardware Architecture of the SECP256K1}
\label{sec:hw_arch}
 SECP256K1 executes modular arithmetic operations, including addition, subtraction, multiplication, and division in an affine coordinate system, i.e., GF($\mathbb{F}_p$) where $ \mathbb{F}_p \in (x,y)$ \cite{panchbhai2015implementation}. However, modular inversion/division is the most expensive in complexity, area, and execution time \cite{guo2023efficient, hossain2015high}. Nevertheless, transforming the coordinates from affine to projective reduces the number of modular division operations performed by SECP256K1 (i.e., GF($\mathbb{F}_p$) where $ \mathbb{F}_p \in (x,y,z)$) \cite{panchbhai2015implementation}. Therefore, \autoref{alg:eqs} depicts a set of equations used to compute the complete \gls{pa} in the projective coordinate system over prime-order elliptic curves. The equations thwart \gls{sca} attacks by removing the branching inherent to the addition operation of the short Weierstrass \glspl{ec}. The branching causes timing variability that could leak secret data  \cite{renes2016complete}. This work employs the equations inside the Montgomery Larder algorithm to perform SECP256K1 \gls{pm} in projective coordinates. 
 
 However, SECP256K1 must perform one final modular division to return the final results to affine coordinates, i.e $(x, y, z) \Rightarrow (xz^{-1}, yz^{-1})$. SECP256K1 utilizes the \gls{bia} to compute the modular division. The algorithm is based on the \gls{eea} which calculates the multiplicative inverse of an integer $z \in \mathbb{F}_p$ by calculating two variables $R$ and $q$ that satisfy $zR + pq = \text{gcd}(z, p) = 1$, where gcd is a function used to calculate the greatest common divisor of two numbers \cite{hossain2015high}. Hence, the proposed architecture comprises two main parts, SECP256K1 \gls{pm} and \gls{bia}.

 \subsection{Architecture of SECP256K1 ECPM}
\gls{pm} is accomplished by employing \gls{pa} and \gls{pd} in the Montgomery Ladder algorithm as shown in \autoref{alg:mont_orig}. Moreover, \gls{pd} is computed using \gls{pa} where the two points are similar (i.e., $\bm{R} = \bm{P} + \bm{P}$ = 2$\bm{P}$). Therefore, we commence by designing the \gls{pa} module. This work uses the equations in \autoref{alg:eqs} to design \gls{pa}.

All operations in \autoref{alg:eqs} are modulo $\texttt{P}$, where $\texttt{P}$ is a specific prime number chosen for SECP256K1 shown in \autoref{table:tab_secp}. Therefore, we first design a \gls{malu} to perform modular addition, subtraction, and multiplication. A shift-and-add algorithm is utilized to perform the modular operations \cite{opencores0}. Subsequently, the proposed Montgomery Ladder algorithm employing the \gls{pm} architecture is shown in \autoref{alg:montg}. A temporary register file $\bm{R_t}$ is added to prevent variability and maintain uniformity in both branches when processing $k_i$ = 1 and $k_i$ = 0. $\bm{R_t}$ is loaded with $\bm{R_0}$ when $k_i$ = 1 and $\bm{R_1}$ when $k_i$ = 0. Moreover, registers $\bm{R_0}$, $\bm{R_1}$, and $\bm{R_t}$ are loaded in parallel. This helps prevent further variations where a uniform control structure that does not depend on the order of operations is created \cite{kabin2020breaking}.

\begin{figure}
\centering
    \resizebox{0.45\textwidth}{!}{\input{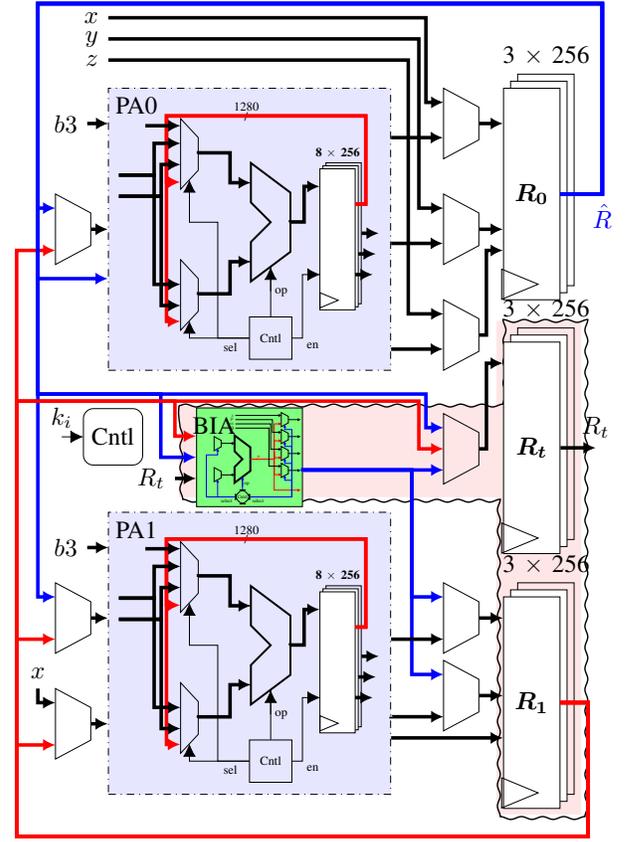}}
    \caption{The proposed hardware architecture of SECP256K1 hash function.}
    \label{fig:eth_pa}
\end{figure}

\autoref{fig:eth_pa} shows the proposed hardware architecture for the Montgomery Ladder algorithm in \autoref{alg:montg} implementing the SECP256K1 hash function. Moreover, the doted blue modules are the \gls{pa} hardware architecture implementing the equations in \autoref{alg:eqs}. The Montgomery Ladder architecture utilizes two \gls{pa} modules that run in parallel. A \gls{cntl} takes the bit values $k_i$ of the private key and controls all the select and enable signals for multiplexers and registers respectively. Moreover, the solid green module shows the \acrfull{bia} architecture.


  \begin{algorithm}
    \caption{Montgomery Ladder algorithm with temporary registers. Adapted from\cite{montgomery1987speeding}}
    \label{alg:montg}
    \begin{algorithmic}[1]\small
    \renewcommand{\algorithmicrequire}{\textbf{Input:}}
    \renewcommand{\algorithmicensure}{\textbf{Output:}}
    \Require $\bm{P} \in (x, y, z), \bm{k}=(k_{t-1}, \cdots, k_0)$ with $k_{t-1}=1$
     \State $\bm{R_0} \leftarrow \bm{P}; \bm{R_1} \leftarrow 2\bm{P}$ 
     \For {$i = t-2$ : $0$}
        \If {$k_i = 1$} $\bm{R_0} \leftarrow \bm{R_0} + \bm{R_1}; \bm{R_1} \leftarrow 2\bm{R_1}; \textcolor{blue}{\bm{R_t} \leftarrow 2\bm{R_0}}$\label{montg:if-st-r0}
        \Else \hspace{4.4em} $\bm{R_1} \leftarrow \bm{R_0} + \bm{R_1}; \bm{R_0} \leftarrow 2\bm{R_0}; \textcolor{blue}{\bm{R_t} \leftarrow 2\bm{R_1}}$\label{montg:if-else-r1}
     \EndIf
     \EndFor \label{montg:endfor}
     \Ensure  $\hat{\bm{R}}=\bm{nP}$
    \end{algorithmic} 
 \end{algorithm}

\begin{table*}[]
    \centering
    \caption{Comparing implementation results of the SECP256K1 Algorithm.}
    \begin{threeparttable}
    \begin{tabular}{l l l l l l l l l l l}\hline
        Work & Platform & \multicolumn{4}{c}{Area} & Frequency & \multicolumn{2}{c}{Latency}  & Throughput\tnote{a} \\
        & & kLUT & DSP & RAM (kbits) & Registers & (MHz) & (ms) & (kCC)& (kbps)\\
        \hline
         \textbf{This work} & \textbf{Zynq-US} & \textbf{21} & \textbf{0} & \textbf{0} & \textbf{13\,881} & \textbf{250} & \textbf{7.58} & \textbf{1\,895} & \textbf{34}\\
         \textbf{This work} & \textbf{Artix-7} & \textbf{24} & \textbf{0} & \textbf{0} & \textbf{13\,385} & \textbf{90} & \textbf{21} & \textbf{1\,895} & \textbf{12}\\
         Mehrabi et al.\cite{mehrabi2020elliptic} & Virtex-7 &  47 & 560 & 0 & 29\,742 & 125 & 0.25& N/A & N/A\\
         Asif et al.\cite{asif2018fully} & Virtex-7 & 19 & 1\,036 & 828 & N/A & 87 & 0.73 & 63 & 351\\
         Islam et al.\cite{islam2019fpga} & Virtex-7 & 36 & N/A & N/A & N/A & 178 & 1.48 & 2630& 173\\
         Romel et al.\cite{romel2023fpga} & Virtex-7 & 52 & 0 & N/A & 15\,263 & 122 & 0.54 & 66& 476\\
         Arunachalam et al.\cite{arunachalam2022fpga} & Virtex-5 & 33 & N/A & N/A & N/A & 192 & 1.21 & 232 & 212\\
         Roy et al.\cite{roy2012implementation} & Virtex-5 & 40 & 0 & N/A & N/A & 43 & 0.60 & 26& 1\,667\\
         Asif et al.\cite{asif2017high} & Virtex-7 & 97 & 2799 & 7\,452 & N/A & 73 & 2.96 & 216 & 1\,816\\
        \hline 
    \end{tabular}
     \begin{tablenotes}
            \item[a] Throughput is estimated by authors as (Frequency $\div$ CC) $\times$ 256.
        \end{tablenotes}
    \end{threeparttable}
    \label{table:secp}
\end{table*}

\subsection{Architecture of Binary Inversion}
\label{subsec:bin_inv}
The output of the Montgomery Ladder $\hat{\bm{R}}$ in \autoref{fig:eth_pa} is in projective coordinate form (i.e., $\bm{\hat{R}}$ = (x$_0$, y$_0$ z$_0$) $\in$ GF(\texttt{p})). Therefore, $\bm{\hat{R}}$ must be converted to affine coordinates to represent a public key in a \gls{cryp} wallet. The point in affine coordinates is computed by calculating the modular multiplicative inverse of the value in the z-axis and multiplying it with the x- and y-axis values of $\bm{\hat{R}}$, i.e, (x$_0\bm{r}$, y$_0\bm{r}$) $\Rightarrow$ (x$_1$, y$_1$), where $\bm{r}$ = $z^{-1}$ mod $\texttt{p}$. This work employs the \acrfull{bia} to calculate $\bm{r}$ \cite{hossain2015high}. The algorithm uses additions, subtractions, and shifting operations to compute the multiplicative inverse. Due to its expensive resource consumption, this is the only modular division performed in SECP256K1 implementation using projective coordinates. The solid green module in \autoref{fig:eth_pa} depicts the proposed architecture of \gls{bia}. The architecture also reuses registers $\bm{R_1}$ and $\bm{R_t}$ shown by the rugged red region. After execution, the affine coordinates (i.e, (x, y)) are stored in $\bm{R_t}$. \Gls{cryp} wallets use the coordinate as a public key K$_{pub}$. 


\section{FPGA Implementation Results and Discussion}
\label{sec:fpga_imp}
A Xilinx ZCU104 and an Artix-7 \gls{fpga} boards were used to implement the proposed SECP256K1 architecture. Moreover, Vivado 2022.2 was used for simulation and a reference software implementation was used to verify the output. The comparison in \autoref{table:secp} evaluates the proposed implementation against state-of-the-art solutions. The area metric comprises \glspl{lut}, \gls{dsp} blocks, \gls{ram} blocks, and registers. Moreover, latency is measured in milliseconds (ms) and \gls{cc}. Throughput is estimated as (Frequency $\div$ CC) $\times$ $k$, where $k$ is the size of the output in bits \cite{romel2023fpga}. 

Our implementation stands out for its minimal \gls{lut} count, surpassed only by \cite{asif2018fully}. However, \cite{asif2018fully} requires significantly more \glspl{dsp} and \gls{ram} blocks, highlighting a critical efficiency trade-off. Unlike other designs such as \cite{romel2023fpga} and \cite{roy2012implementation}—which also avoid \glspl{dsp} but demand higher \glspl{lut}—our approach efficiently manages operations with fewer resources. The lack of \gls{ram} and minimal register usage in our design further underscores its suitability for resource-constrained, low-power applications like \gls{cryp} wallets, suggesting it could offer enhanced scalability and thermal performance compared to comparable architectures.

\section{Conclusion}
\label{sec:concl}

The proposed hardware architecture for the SECP256K1 algorithm enhances resistance against \gls{sca} attacks by maintaining uniformity in register operations during the private key processing. Moreover, the architecture is designed for \gls{cryp} wallet application, where it utilizes minimum resources and adheres to the industry standard for small, portable \gls{cryp} wallets. Implementation results indicate that the proposed architecture requires, on average, 45\% less \glspl{lut} compared to analogous implementations in literature. Future work will focus on performing detailed experiments on the device to demonstrate the security against \gls{sca} attacks and integrate the architecture into a \gls{cryp} hardware wallet.

\bibliographystyle{unsrt}
\bibliography{references}

\end{document}